\documentclass[3p,twocolumn]{elsarticle}
\usepackage{graphicx}
\usepackage{amssymb}
\usepackage{amsmath}

\usepackage{algorithmic}

\journal{Computer Physics Communications}

\begin{document}

\begin{frontmatter}

\title{Duality of critical interfaces in Potts model: numerical check}
\address[landau]{Landau Institute for Theoretical Physics, 142432
  Chernogolovka, Russia}
\address[mipt]{Moscow Institute for Physics and Technology, 141700
  Dolgoprudny, Russia}

\author[landau,mipt]{Alexey Zatelepin}
\author[landau]{Lev Shchur}

\begin{abstract}
  We report on numerical investigation of fractal properties of
  critical interfaces in two-dimensional Potts models. Algorithms for finding
  percolating interfaces of Fortuin--Kasteleyn clusters, their
  external perimeters and interfaces of spin clusters are
  presented. Fractal dimensions are measured and compared to exact
  theoretical predictions.
\end{abstract}
\end{frontmatter}

\section{Introduction}
Study of the fractal geometry of critical interfaces has
recently become one of the main areas of development in statistical
physics. Indeed, the SLE (Schramm--L\"owner Evolution) process discovered by
O.~Schramm~\cite{schramm00} is defined in terms of fractal curves. This
discovery has led to an explosion of rigorous results for statistical
models on a plane (see for example review~\cite{bauer06}).

Potts model~\cite{potts} is a generalization of Ising
model where spin variables $\sigma_i$ can possess values in $\{1, 2,
\ldots, q\}$. Hamiltonian of the model with nearest--neighbor
interactions is
\begin{equation}
  \label{eq:hamilt}
  \mathcal{H} = -\sum_{<ij>}J_{ij}(\delta_{\sigma_i\sigma_j} - 1).
\end{equation}

We consider ferromagnetic model with no bond disorder,
that is $J_{ij}{=}J{>}0$. It is convenient to rewrite partition
function as a sum over bond variable configurations $\{b_{ij}\}$.
\begin{equation}
  \label{eq:fk}
  \mathcal{Z_{\mathrm{Potts}}} = \sum_{\{b_{ij}\}} \left(\prod_{b_{ij} = 1}p\right) \left(\prod_{b_{ij} = 0}(1-p)\right)q^{\#c},
\end{equation}
where $\#c$ is the total number of clusters in the configuration and
$p {=} 1 - {\exp(-\beta J)}$. This representation is called
Fortuin--Kasteleyn representation~\cite{fk}. Note that now $q$ is not
restricted to integer values. Two-dimensional (2D) Potts model has a
second order phase transition for $q\in [0; 4]$. For greater values
of $q$ phase transition becomes first--order.

Correspondence between critical Potts models and SLE remains
hypothetical (rigorously proved by Smirnov only for the case of
square--lattice Ising model~\cite{smirnov06}, for numerical
check see~\cite{gliozzi10}). For these models SLE
approach is complementary to older and less rigorous Coulomb
gas~\cite{Nienhuis} and Conformal Field Theory~\cite{CFT}
methods. 
Elegant prediction by
Duplantier~\cite{duplantier00} (confirmed by SLE, provided there is
correspondence) states that the fractal dimensions of boundaries of
$d_{FK}$ and their external perimeters $d_{EP}$ satisfy duality
relation $(d_{FK} - 1)(d_{EP} - 1) = 1/4$. This prediction is
supported by numerical results presented by Aisikainen et
al~\cite{aisikainen03} and by more recent data of Adams et
al~\cite{adams10}.  For boundaries of spin clusters of Ising model
Smirnov's paper predicts that their fractal dimension is equal to
$d_{EP}$. The hypothesis~\cite{gamsa05} that it is true for other
values of parameter $q$ is supported by recent numerical data of
Jacobsen et al~\cite{jacobsen09} for $q=3$.

We present algorithm for identification of percolating boundary of
Fortuin--Kasteleyn (FK) clusters, of percolating external perimeter of
FK clusters (EP) and percolating boundary of spin clusters (SP).
Preliminary estimations of fractal dimension of critical FK interfaces,
EP, and SP interfaces for various values of $q\in[1;4]$ are presented.

In section 2 we review Monte--Carlo algorithms. In section 3 we
proceed with the definitions of clusters, their boundaries. In section
4 we introduce algorithms for finding lengths of interfaces. We
conclude with the account of our simulations.

\section{Monte--Carlo algorithms}

Well known algorithms for Monte--Carlo simulations of Potts model with
noninteger $q$ include Chayes--Machta algorithm~\cite{chayes98} and Sweeny
algorithm~\cite{sweeny83}.

Sweeny algorithm is essentially a Metropolis type algorithm for
simulating partition function~(\ref{eq:fk}) via single--bond
updates. It is applicable for simulations in the whole range $q\in [0,
4]$. As this algorithm needs nonlocal information for local updates,
its performance depends strongly on the nature of implementation. The
original article describes some optimizations for 2D lattice which
help achieve $O(N\log N)$ performance asymptotic (where $N = L^2$ is
the number of lattice sites).

Chayes--Machta algorithm is easy to implement, sweeps the whole
lattice in $O(N)$ steps and works for $q\in[1; 4]$. One step of the
algorithm proceeds as follows:
\begin{enumerate}
\item Find all bond clusters in the configuration.
\item Independently label clusters as ``active'' (site color 1) with
  probability $1/q$ or ``inactive'' (site color 0) with probability
  $(q-1)/q$.
\item Erase all bonds. Independently with probability $p$ add bonds
  between sites of active clusters.
\end{enumerate}

There is some freedom in which cluster--finding algorithm to use. We
used Newman--Ziff method~\cite{newmanziff} as it turned out to be
slightly superior in performance to other methods (e.g. breadth--first
search).

As Chayes--Machta algorithm (when it is applicable) generally performs
better than Sweeny algorithm, we used it for all our
simulations. Additionally, as will be shown later, this algorithm
provides an idea for the identification of spin cluster applicable to
noninteger values of $q$.

\section{Definition of clusters and their interfaces}

Our goal in this section is to provide unified definitions for all the
objects considered. Let us start with the definition of clusters.

\subsection{FK and spin clusters}

Given lattice graph $G$ and bond configuration $b \equiv \{b_{ij}\}$
on it, we call every group of vertices of $G$ which is a connected
component of $b$ a \emph{cluster} of $b$. This definition allows us to
treat clusters of different types merely as clusters on different bond
configurations.

Specifically, let bond configuration $b$ be drawn from the
distribution induced by the Fortuin--Kasteleyn partition
function~(\ref{eq:fk}). Then, clusters of $b$ are called
\emph{Fortuin--Kasteleyn} or FK clusters.

\emph{Spin} clusters are defined in the following way: let
${\sigma_i}$ be the configuration of spin variables obtained with
Chayes--Machta (CM) algorithm. Then, generate bond configuration by setting
all $b_{ij} {=} 1$ between sites $i, j$ such that
$\sigma_i{=}\sigma_j{=1}$ and all other $b_{ij} {=} 0$. Note that such
procedure is equivalent to performing the bond--adding step of CM
algorithm at zero temperature.

When $q$ is integer, ``active'' sites in CM algorithm represent sites
of one particular Potts color. So for integer $q$ this definition
coincides with the regular one and provides natural extension to
noninteger values of $q$.

\subsection{Cluster boundary}

Any bond configuration induces a configuration of closed
loops~\cite{bkw} on a medial lattice (sites of the medial lattice are
in the middles of the bonds of the original lattice), which is defined
for any planar lattice. Let us define \emph{cluster boundary} as the
set of loops on a medial lattice adjacent to at least one of the sites
of the cluster considered.

\begin{figure}
  \begin{center}
      \resizebox{45mm}{!}{\includegraphics{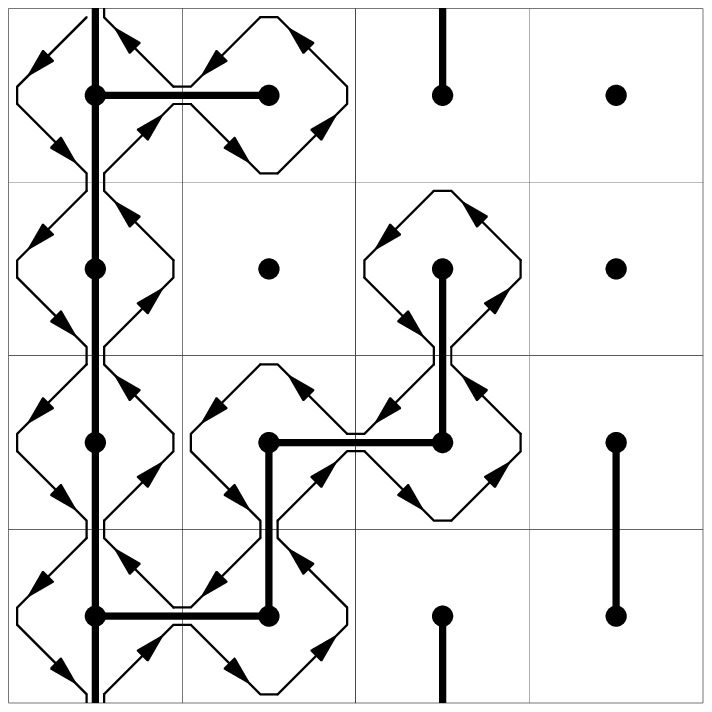}}
      \caption{Example of bond conguration with
nontrivial part of the boundary of a cluster marked by the line with
arrows.}
    \label{fig:iface}
  \end{center}
\end{figure}

On a lattice with periodic boundary conditions (as is indeed the case
in our simulations) most of these loops can be contracted into a point
by a continuous transformation. But for loops that wind nontrivially
on the torus of the lattice (we will call such loops ``nontrivial'')
this is not true. We will be interested primarily in nontrivial loops,
as they possess an unambiguous length scale (namely, lattice size $L$)
and do not disappear in the scaling limit as $L{\to} \infty$. Example
of a cluster with nontrivial part of the boundary marked is presented
in Fig.~\ref{fig:iface}.

\subsection{External perimeter}

\begin{figure}
  \begin{center}
      \resizebox{45mm}{!}{\includegraphics{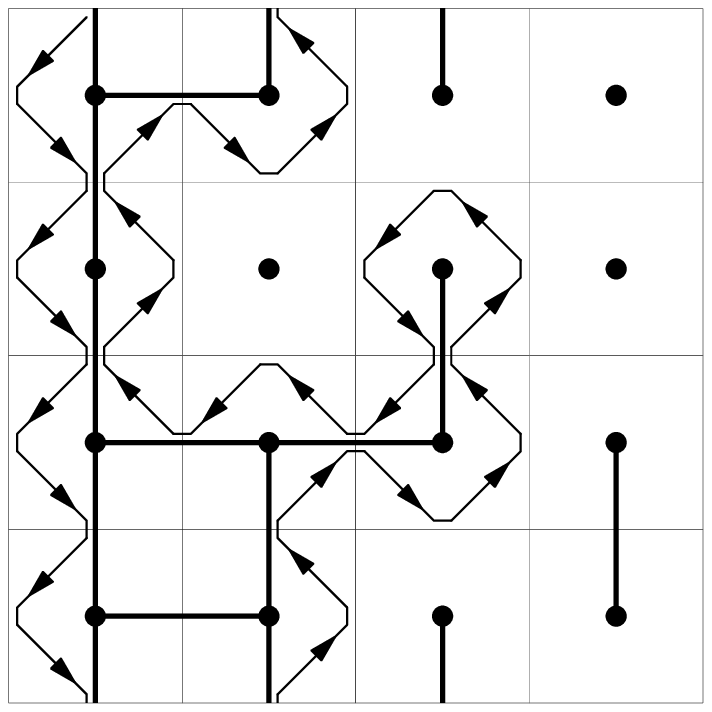}}
    \caption{External perimeter.}
    \label{fig:iface_ep}
  \end{center}
\end{figure}

For any cluster $c$ with a nontrivial boundary let us set all bond variables
$b_{ij} {=} 1$ for all neighboring sites $i, j {\in} c$ such that
there is a loop adjacent to both $i$ and $j$. As a result the initial
cluster boundary is split into trivial loops (which encircle fjords
that become ``lakes'') and two nontrivial parts. We will call them the
\emph{external perimeter} of a cluster. External perimeter for the
cluster in Fig.~\ref{fig:iface} is presented on
Fig.~\ref{fig:iface_ep}.

The definition makes use of the fact that the loop is ``internal''
iff it is trivial. For other types of boundary condition one has to
come up with some other method for distinguishing between internal and
external loops.

According to this definition only the fjords with the narrowest
entrances (the ones one lattice spacing wide) are closed. Our
simulations suggest that it is sufficient to achieve the expected fractal
dimension, although one could in principle define the whole hierarchy
of external perimeters as was found in~\cite{aharony87}.

Note also that according to presented definition external perimeter of a
spin cluster coincides with its regular outer boundary.

\section{Algorithms for tracing boundaries}

\begin{table*}[!t]
\begin{center}
  \begin{tabular}{llllllllll}
    \hline
    $q$ & \quad & $d_{FK}^{th}$ & measured & \qquad & $d_{EP}^{th}$ &
    measured & \qquad & $d_{spin}^{th}$ & measured \\
    \hline
    1 && 1.75 & 1.75002(2) && 4/3 & 1.33331(9) && -- & -- \\
    1.5 && 1.70444 & 1.70449(7) && 1.35489 & 1.3546(2) && 1.35489 & 1.3549(2) \\
    2 && 5/3 & 1.6667(1) && 1.375 & 1.3747(5) && 1.375 & 1.37514(8) \\
    2.5 && 1.63274 & 1.63275(8) && 1.39511 & 1.3953(9) && 1.39511 & 1.3946(3) \\
    3 && 1.6 & 1.6002(1) && 1.41667 & 1.418(1) && 1.41667 & 1.418(1) \\
    3.5 && 1.56498 & 1.565(2) && 1.44248 & 1.438(4) && 1.44248 & 1.46(2) \\
    4 && 1.5 & 1.534(1) && 1.5 & 1.405(1) && 1.5 & 1.428(1) \\
    \hline
  \end{tabular}
\end{center}
\caption{Fractal dimensions: theoretical predictions and measured values.}
\label{tab:results}
\end{table*}

The main purpose of the algorithms in this section is to find lengths
of cluster boundaries and their external perimeters given the
configuration $(\{b_{ij}\},\{s_i\})$. In the basis of all these
algorithms lies a lattice walker which traces loops on a medial
lattice, with one segment at a time. We will present definitions for a
square lattice as generalizations to arbitrary planar lattice are
straightforward.

\begin{figure}
  \begin{center}
      \resizebox{45mm}{!}{\includegraphics{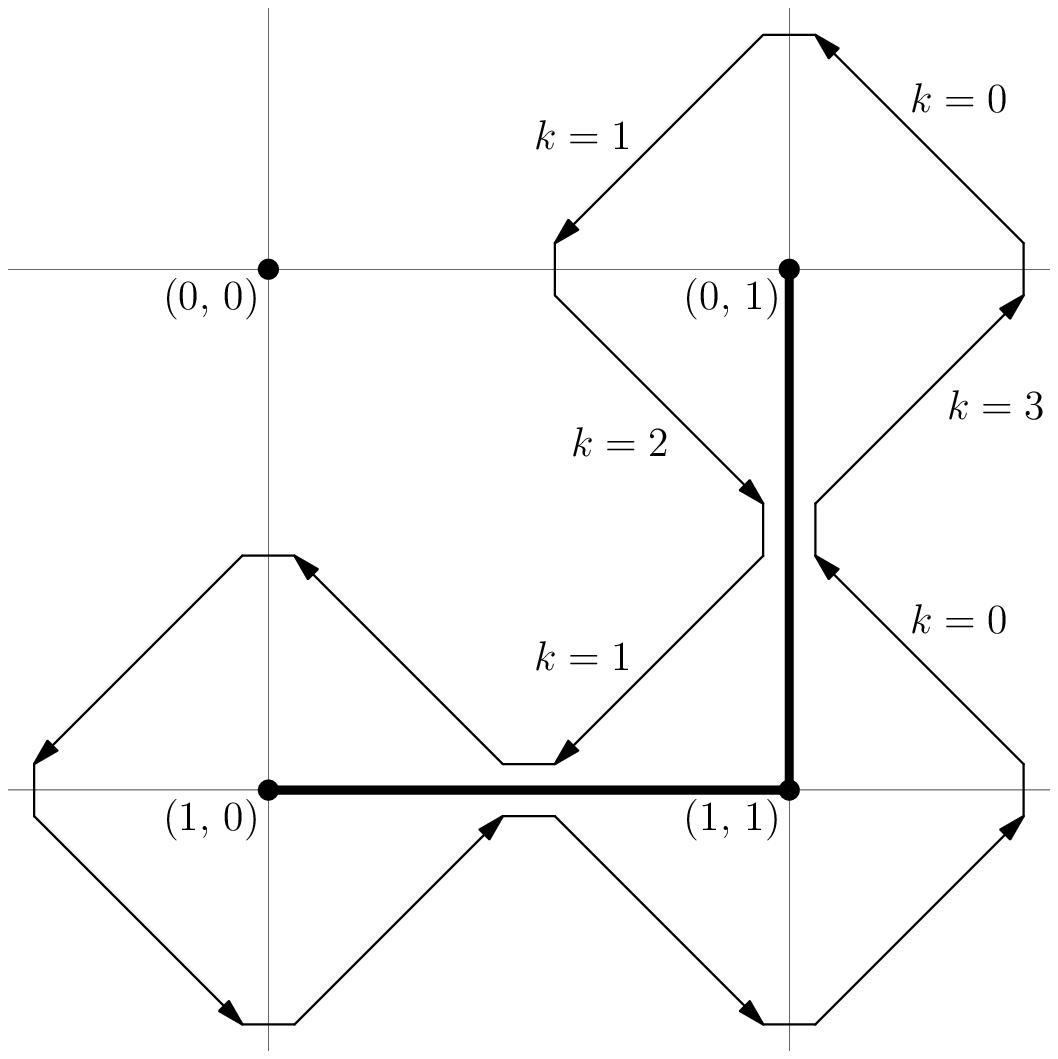}}
    \caption{Tracing the boundary.}
    \label{fig:nl}
  \end{center}
\end{figure}

The state of a walker is a tuple $e = (i, j, k)$ representing the
segment of a medial lattice. Here $(i, j)$ are the lattice coordinates
and $k \in \{0, 1,2, 3\}$ is the index of the segment (see
Fig.~\ref{fig:nl}). Procedure $\mathtt{NextSegment}(e)$ finds next to
$e$ segment in a loop to which $e$ belongs. If the corresponding bond
ending in $(i, j)$ is not present, $\mathtt{NextSegment}(i, j, k) =
(i, j, (k + 1) \% 4)$, where $\%$ represents taking a remainder. If
the bond is present, then e.g. $\mathtt{NextSegment}(i, j, 2) = ((i +
1)\% L, j, 1)$. For other cases $\mathtt{NextSegment}$ is defined
analogously. Note that to trace spin cluster boundaries instead of FK
one does not need to construct another configuration but only tell the
walker that the bond is present if respective sites are both active.

Procedure $\mathtt{TraceLoop}(e)$ (Fig.~\ref{traceloop}) returns the
tuple $(l, wx, wy)$, where $l$ is the length the loop containing $e$
and $wx, wy$ are the number of times the loop winds around the lattice
in both directions. $\mathtt{Windings}(e)$ returns $(0,0)$ for a
segment in the bulk and one of $\{(1,0), (-1, 0), (0, 1), (0, -1)\}$
if the walker needs to cross the edge of the lattice in a corresponding
direction. The loop is nontrivial iff $(wx, wy) \ne (0,0)$ (note that
the mere fact that the walker crosses the lattice edge is
insufficient).

\begin{figure}
\begin{algorithmic}
  \STATE $l \leftarrow 0, wx \leftarrow 0, wy \leftarrow 0$
  \STATE $e' \leftarrow e$
  \REPEAT
  \STATE $l \leftarrow l + 1$
  \STATE $(wx, wy) \leftarrow (wx, wy) + \mathtt{Windings}(e)$
  \STATE $e' \leftarrow \mathtt{NextLink}(e')$
  \UNTIL{$e' \ne e$}
  \RETURN{$(l, wx, wy)$}
\end{algorithmic}
\caption{$\mathtt{TraceLoop}(e)$}
\label{traceloop}
\end{figure}

Procedure $\mathtt{FindNontrivialLoops}()$ (Fig.~\ref{nontrivialloops})
finds the set of all nontrivial loops in the configuration. The subset
of boundary segments
\begin{multline}
  \label{eq:exits}
  \mathtt{Exits} = \{(i,j,k):i{=}0, 0{<}j{<}L, k{\in}\{0,1\}\} \cup\\
  \{(i,j,k): 0{<}i{<}L, j{=}0, k{\in}\{1,2\}\}
\end{multline}
is chosen in such way that any nontrivial loop has a segment in this
subset. A naive way to find nontrivial loops would be to launch
$\mathtt{TraceLoop}$ for all segments in $\mathtt{Exits}$. But, as
one loop can cross $\mathtt{Exits}$ many times, this leads to much
worse than $O(N)$ performance. To avoid this we mark all already
traced loops with an unique label. $\mathtt{ClearLabels}$ sets all
the labels to zero.

\begin{figure}
\begin{algorithmic}
  \STATE $\mathtt{ClearLabels}()$
  \STATE $loops \leftarrow \emptyset$
  \STATE $label \leftarrow 1$
  \FOR {$e \in \mathtt{Exits}$}
  \IF {$\mathtt{Label}(e) = 0$}
  \STATE $(l, wx, wy) \leftarrow \mathtt{TraceLoop}(e)$
  \STATE $\mathtt{LabelLoop}(e, label)$
  \STATE $label \leftarrow label + 1$
  \IF {$wx \ne 0$ \OR  $wy \ne 0$}
  \STATE $loops \leftarrow loops \cup e$
  \ENDIF
  \ENDIF
  \ENDFOR
  \RETURN $loops$
\end{algorithmic}
\caption{$\mathtt{FindNontrivialLoops}()$}
\label{nontrivialloops}
\end{figure}

Search for external perimeters is performed only for a configuration
with nonempty $loops$ set. For all $e{\in}loops$ define the set
\begin{multline}
  \label{eq:loopexits}
  \mathtt{LoopExits}(e) =\\ \mathtt{Exits} \cap \{e': e' =
  \mathtt{NextSegment}^n(e)\}
\end{multline}
of all segments of loop represented by $e$ which are also in
$\mathtt{Exits}$. We will also need procedures
$\mathtt{NextEPSegment}(e)$ and $\mathtt{TraceEP}(e)$ defined
analogously to $\mathtt{NextSegment}$ and $\mathtt{TraceLoop}$ only in
such way that bonds added according to the definition of external
perimeter are taken into account (again, no modification of
configuration is necessary). The fact that the loop segments are still
labeled after the run of $\mathtt{FindNontrivialLoops}$ greatly helps
in recognizing gates to the fjords.

Now finding the length of the external perimeter is easy
(Fig.~\ref{eplengths}).
\begin{figure}
\begin{algorithmic}
  \STATE $lengths \leftarrow \emptyset$
  \FOR {$e \in loops$}
  \FOR {$e' \in \mathtt{LoopExits}(e)$}
  \STATE $(l, wx, wy) \leftarrow \mathtt{TraceEP}(e')$
  \IF {$wx \ne 0$ \OR  $wy \ne 0$}
  \STATE $lengths \leftarrow lengths \cup l$
  \ENDIF
  \ENDFOR
  \ENDFOR
  \RETURN $lengths$
\end{algorithmic}
\caption{$\mathtt{FindEPLengths}(loops)$}
\label{eplengths}
\end{figure}

Processing a configuration with these algorithms requires $O(N)$ steps.

\section{Simulation and results}

We simulated 2D Potts model on a square lattice with periodic boundary
conditions and $q\in\{1,1.5,2,2.5,3,3.5,4\}$. For pseudorandom number
generation Mersenne Twister algorithm~\cite{mersenne} (MT19937) was
used. For each value of $q$ lattices of linear size $L$ from 32 to
1323 were simulated. For each lattice size we performed 150 to 1000
independent runs. One independent run started with a thermalization
period of $2 \cdot 10^4$ Monte--Carlo steps. It was followed by $10^5$
steps during each of which the configuration was searched for
nontrivial interfaces. After each run mean values of the lengths of
interfaces of all types were computed. Final output of the simulation
is the mean values of interface lengths for all values of $q$ and $L$.

Fractal dimensions were extracted using least--squares
fitting. To take corrections to scaling into account, data was fitted
using the following functions:
\begin{eqnarray}
  l &\approx& AL^{d_f}(1+b/L), \label{eq:linfit}\\
  l &\approx& AL^{d_f}(1+b/L^c), \label{eq:nlinfit}
\end{eqnarray}
which correspond to analytical and non--analytical main correction
term respectively. We tried to fit with analytical correction first,
discarding data points with small $L$ until satisfactory $\chi^2$ per
degree of freedom was obtained. If this procedure was unsuccessful,
function~(\ref{eq:nlinfit}) was used.

Results are presented in Table~\ref{tab:results}. Except $q=4$, they
are in close agreement with theoretical predictions. The reason for
the discrepancy at $q=4$ is that main correction term becomes
logarithmic, which is very difficult to properly take into account.

\bibliographystyle{elsarticle-num}
\bibliography{sle,numeric}

\begin{thebibliography}{10}
\expandafter\ifx\csname url\endcsname\relax
  \def\url#1{\texttt{#1}}\fi
\expandafter\ifx\csname urlprefix\endcsname\relax\def\urlprefix{URL }\fi
\expandafter\ifx\csname href\endcsname\relax
  \def\href#1#2{#2} \def\path#1{#1}\fi

\bibitem{schramm00}
O.~Schramm, Scaling limits of loop--erased random walks and uniform spanning
  trees, Isr. J. Math. 118 (2000) 221.

\bibitem{bauer06}
M.~Bauer, D.~Bernard, 2d growth processes: Sle and loewner chains, Phys. Rep.
  432~(3-4) (2006) 115.

\bibitem{potts}
R.~B. Potts, Some generalized order-disorder transformations, Proc. Camb. Phil.
  Soc. 48 (1952) 106.

\bibitem{fk}
C.~M. Fortuin, P.~W. Kasteleyn, On the random-cluster model: I. introduction
  and relation to other models, Physica (Amsterdam) 57 (1972) 536.

\bibitem{smirnov06}
S.~Smirnov, Towards conformal invariance of 2d lattice models, Proceedings of
  the International Congress of Mathematicians (ICM), Madrid, Spain, August
  22-30, 2006 Vol. II (2006) 1421.

\bibitem{Nienhuis}
B.~Nienhuis, Coulomb gas formulation of two--dimensional phase transitions, in:
  Phase Transitions and Critical Phenomena, Vol.~11, Academic Press, 1987, pp.
  1--54.

\bibitem{CFT}
A.~A. Belavin, A.~M. Polyakov, A.~B. Zamolodchikov, Infinite conformal symmetry
  in two-dimensional quantum field theory, Nucl. Phys. B 241 (1984) 333.

\bibitem{gliozzi10}
F.~Gliozzi, M.~A. Rajabpour, {Conformal Curves in Potts Model: Numerical
  Calculation}, J. Stat. Mech. 5 (2010) L05004.

\bibitem{duplantier00}
B.~Duplantier, Conformally invariant fractals and potential theory, Phys. Rev.
  Lett. 84 (2000) 1363.

\bibitem{aisikainen03}
J.~{Asikainen}, A.~{Aharony}, B.~B. {Mandelbrot}, E.~{Rausch}, J.~{Hovi},
  {Fractal geometry of critical Potts clusters}, Eur. Phys. J. B 34 (2003)
  479--487.

\bibitem{adams10}
D.~A. {Adams}, L.~M. {Sander}, R.~M. {Ziff}, {Fractal dimensions of the Q-state
  Potts model for complete and external hulls}, J. Stat. Mech. 3 (2010) P03004.

\bibitem{gamsa05}
A.~Gamsa, J.~Cardy, Schramm loewner evolution in the three-state potts
  model---a numerical study, J. Stat. Mech. 8 (2007) 20.

\bibitem{jacobsen09}
J.~L. Jacobsen, P.~Le~Doussal, M.~Picco, R.~Santachiara, K.~J. Wiese, Critical
  interfaces in the random-bond potts model, Phys. Rev. Lett. 102~(7) (2009)
  070601.

\bibitem{chayes98}
L.~Chayes, J.~Machta, Graphical representations and cluster algorithms ii,
  Physica A 254 (1998) 477--516.

\bibitem{sweeny83}
M.~Sweeny, Monte carlo study of weighted percolation clusters relevant to the
  potts models, Phys. Rev. B 27 (1983) 4445--4455.

\bibitem{newmanziff}
M.~E.~J. Newman, R.~M. Ziff, Fast monte carlo algorithm for site or bond
  percolation, Phys. Rev. E 64 (2001) 016706.

\bibitem{bkw}
R.~J. Baxter, S.~B. Kelland, F.~Y. Wu, Equivalence of the potts model or
  whitney polynomial with an ice-type model, J. Phys. A 9~(3) (1976) 397.

\bibitem{aharony87}
T.~Grossman, A.~Aharony, Accessible external perimeters of percolation
  clusters, J. Phys. A 20~(17) (1987) L1193.

\bibitem{mersenne}
M.~Matsumoto, T.~Nishimura, Mersenne twister: a 623-dimensionally
  equidistributed uniform pseudo-random number generator, ACM Trans. Model.
  Comput. Simul. 8~(1) (1998) 3--30.

\end{thebibliography}
\end{document}